\def \Mpc {h^{-1}{\rm Mpc}}

\def \farcm{\hbox{$.\!\!^{\prime}$}}
\def \fdeg{\hbox{$.\!\!^{\circ}$}}

\def \kms {{\rm km~s}^{-1}}
\def \msun {h^{-1} M_\odot}
\def \mlsun {h M_\odot/L_\odot}
\def \beqn {\begin{equation}}
\def \eeqn {\end{equation}}

\def \nczeight {1779}

\def \ncznewe {964} %CAIRNS+15R+CS = new or in press

\def \nkcompe {622}	% within 8 deg
\def \necomp {845}
\def \mlkedge {75}
\documentclass[12pt,preprint]{aastex}
\usepackage{epsfig}
\usepackage{natbib}

\begin{document}
\bibliographystyle{apj}

\title{Infrared Mass-to-Light Profile Throughout the Infall Region of the Coma Cluster}
%\shorttitle{Coma Infall Region}
%\shortauthors{Rines et al.}

\author{K. Rines\altaffilmark{1}, M.J. Geller\altaffilmark{2},
M.J. Kurtz\altaffilmark{1}, A. Diaferio\altaffilmark{3},
T.H. Jarrett\altaffilmark{4}, and J.P. Huchra\altaffilmark{1}}
\email{krines@cfa.harvard.edu}

\altaffiltext{1}{Harvard-Smithsonian Center for Astrophysics, 60 Garden St,
Cambridge, MA 02138 ; krines,  mkurtz, huchra@cfa.harvard.edu}
\altaffiltext{2}{Smithsonian Astrophysical Observatory; mgeller@cfa.harvard.edu}
\altaffiltext{3}{Universit\`a degli Studi di Torino,
Dipartimento di Fisica Generale ``Amedeo Avogadro'', Torino, Italy; diaferio@ph.unito.it}
\altaffiltext{4}{IPAC/Caltech 100-22 Pasadena, CA 91225;
jarrett@ipac.caltech.edu}

\begin{abstract}

Using a redshift survey of \nczeight~galaxies and photometry from the
2-Micron All-Sky Survey (2MASS) covering 200 square degrees, we
calculate independent mass and light profiles for the infall region of
the Coma cluster of galaxies.  The redshift survey is complete to 
$K_s=12.2$ (\nkcompe~galaxies), 1.2 magnitudes fainter than
$M^*_{K_s}$ at the  distance of Coma.  We confirm the mass profile
obtained by Geller, Diaferio, \& Kurtz.  The enclosed mass-to-light
ratio measured in the $K_s$ band is approximately constant to a radius
of $10~\Mpc$, where $M/L_{K_s}= \mlkedge\pm 23\mlsun$, in agreement
with weak lensing results on similar scales.  Within
$2.5\Mpc$, X-ray estimates yield similar mass-to-light ratios
(67$\pm32h$).   The constant enclosed mass-to-light ratio with radius
suggests that K-band light from bright galaxies in clusters traces the
total mass on scales $\lesssim10~\Mpc$.  Uncertainties in the mass
profile imply that the mass-to-light ratio inside $r_{200}$ may be as
much as a factor of 2.5 larger than that outside $r_{200}$.  These
data demonstrate that K-band light is not positively biased with
respect to the mass; we cannot rule out antibias.  These results
imply $\Omega_m = 0.17 \pm 0.05$.  Estimates of possible variations in
$M/L_{K_s}$ with radius suggest that the density parameter is no
smaller than $\Omega_m \approx 0.08$.

\end{abstract}

\keywords{cosmology: observations --- dark matter --- galaxies:
clusters: individual (Coma) --- galaxies: 
kinematics and dynamics --- galaxies: photometry }

\section{Introduction}

The relative distribution of matter and light in the universe is one of
the outstanding problems in astrophysics.  
Clusters of galaxies, the largest gravitationally relaxed objects in
the universe, are important probes of the distribution of 
mass and light. \citet[][]{zwicky1933} first computed the
mass-to-light ratio of the Coma cluster and found that dark matter
dominates the cluster mass.  Recent determinations
yield mass-to-light ratios of $M/L_{B_j}\sim 250 \mlsun$
\citep[][and references therein]{g2000}.  Equating the
mass-to-light ratio in clusters to the global value provides an
estimate of the mass density of the universe; this estimate is subject to
significant systematic error introduced by differences in galaxy populations
between cluster cores and lower density regions \citep{cye97,g2000}.
Numerical simulations suggest that antibias in cluster cores may cause
cluster mass-to-light ratios to exceed the universal value
\citep{kk1999,bahcall2000,benson00}. 
However, there are few measurements of mass-to-light ratios on scales
of $1-10~\Mpc$ \citep[][hereafter
R00]{elt97,small98,kaiserxx,rines2000} to test this conjecture.   

Because clusters are not in equilibrium outside the virial radius,
neither X-ray observations nor Jeans analysis provide secure mass
determinations at these large radii. There are now two methods of
approaching this problem: weak gravitational lensing \citep{kaiserxx} and
kinematics of the infall region \citep[][]{dg97,diaferio1999}. Kaiser
et al.~analyzed the weak lensing signal from a
supercluster at $z \approx 0.4$; the mass-to-light ratio ($M/L_B$=280
$\pm$40 for early-type galaxy light) is constant on scales up to
$6~\Mpc$.  \citet[][hereafter GDK]{gdk99}, applied the kinematic
method of \citet{dg97} to the infall region of the Coma 
cluster.  GDK reproduced the X-ray derived mass profile and
extended direct determinations of the mass profile to a radius of $10~\Mpc$. 
This method has also been applied to the Shapley Supercluster
\citep{rqcm}, A576 (R00), the Fornax cluster \citep{drink},
and A1644 \citep{tustin}.  R00 found an enclosed mass-to-light ratio
of $M/L_R \sim 300 h$ within 4$~\Mpc$.

Here, we calculate the infrared mass-to-light profile to a radius of
$10~\Mpc$ for the Coma cluster using photometry from the Two Micron
All Sky Survey \citep[2MASS,][]{twomass}.  Within a radius of 8\fdeg0,
the redshift survey is complete to $K_s=12.2$ 
(\nkcompe~galaxies), 1.2 magnitudes fainter than $M^*_{K_s}$
at the distance of Coma \citep[][hereafter K01]{twomasslfn}.  Infrared
light is a better tracer of stellar mass than optical light, at least
in late-type galaxies \citep[][]{gpb96}; it is relatively insensitive
to dust extinction and recent star formation.  Despite these
advantages, there are very few measurements of infrared mass-to-light
ratios in clusters \citep{tustin}.  The physical scale at the redshift
of Coma ($cz_\odot=7093~\kms$; $cz_{CMB}=7361~\kms$) is $1^\circ =
1.25~\Mpc$ ($H_0 = 100~h~\kms, \Omega_m = 0.3, \Omega _\Lambda = 0.7$; 
we assume Coma is at rest with respect to the CMB and use $cz_{CMB}$
for all calculations).

\section{Observations}

\subsection{Spectroscopy}

We have collected \nczeight ~redshifts (\ncznewe ~new or in press) within
8\fdeg0 of the center of the Coma cluster 
\citep[collected from ZCAT\footnote{Available at
http://cfa-www.harvard.edu/$\sim$huchra/zcat}, NED\footnote{Available
at http://nedwww.ipac.caltech.edu},][Geller et al.~in
preparation]{vh93, cd96,falco99, sdsscoma,cs2001}.     
We measured new redshifts with FAST, a long-slit spectrograph \citep{fast}
on the 1.5-m Tillinghast telescope of the Fred Lawrence Whipple
Observatory (FLWO).  We selected targets from digitized images of the
POSS I 103aE (red) plates.  
The redshift catalog is complete to $E\approx15.4$ (\necomp~galaxies).  We
later obtained a small number of redshifts ($\sim$20) to complete the
$K_s\leq 12.2$ sample.  Rines et al.~(in preparation)
describes this catalog in 
detail\footnote{The redshift catalog is available at 
http://tdc-www.harvard.edu/comacz/}.

\subsection{2MASS Photometry}

2MASS is an all-sky survey
with uniform, complete photometry \citep{twomasscalib} in three
infrared bands (J, 
H, and K$_s$, a modified version of the K filter truncated at longer
wavelengths). We use a preliminary version of the complete extended
source catalog \citep{twomassxsc}. Future recalibrations may change
the zero-points of individual scans by up to 0.03 mag.  We use the
default $K_s$-band survey magnitudes which include light within the
circular isophote corresponding to $\mu_{K_s}$=20 mag/arcsec$^2$
\citep{twomassxsc}.  These magnitudes omit $\sim$15\% of the flux
(K01).  The sky coverage of the catalog is complete to $K_s$=12.2.
We include 2 galaxies not in 2MASS with $K$ magnitudes from
\citet{gb96}.  There are \nkcompe~galaxies with $K_s\leq 12.2$
within 8\fdeg0 of the center of Coma; all of these galaxies have
measured redshifts.  We make no correction for galactic 
extinction, which is negligible in the near-infrared at the North
Galactic Pole.

\section{Defining the Infall Region with Caustics}

\begin{figure*}
\figurenum{1}
\label{caustics}
%\epsscale{0.4}
\plotone{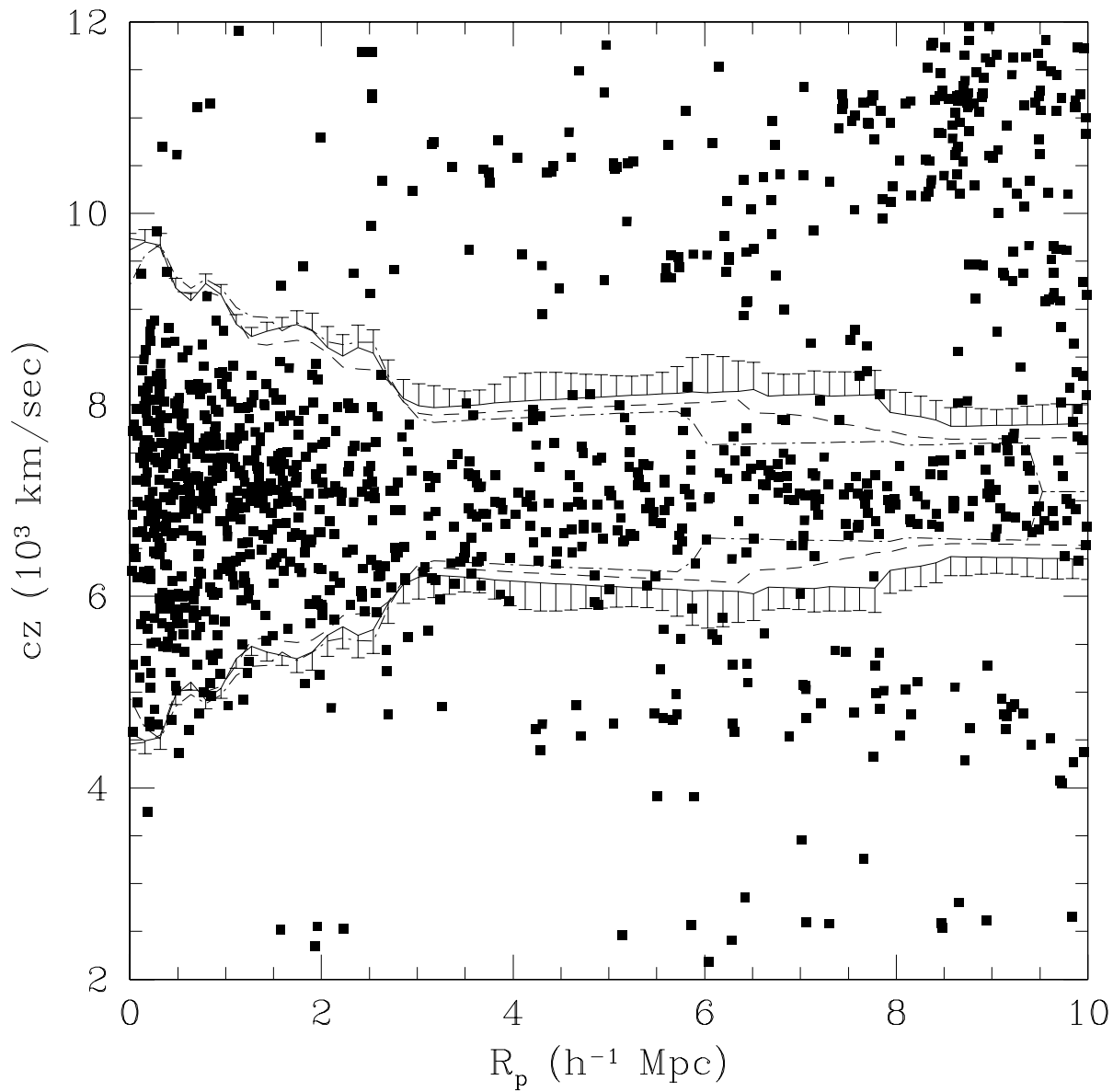} 
\caption{Redshift versus projected radius of galaxies around the Coma
cluster.  The trumpet-shaped caustic pattern which defines the infall
region is clearly visible.  The dashed, solid, and dashed-dotted lines
show the location of the caustics for $q=10$, 25, and 50 with
$1-\sigma$ uncertainties shown for $q=25$. For clarity, the
uncertainties are only displayed away from the cluster.} 
\end{figure*}

Figure \ref{caustics} displays the projected radii and redshifts of
galaxies surrounding Coma.  The expected caustic pattern is easily
visible; we calculate the shape with the technique described in
\citet{diaferio1999} using smoothing parameter $q$ of 10, 25, and 50 to
test the variation caused by the subjective choice of this parameter.
GDK show that the mass profile is robust with respect to limiting
magnitude and non-uniform sampling.
We recalculate the caustics based on additional redshifts collected
since the calculation by GDK and find the same results. The cluster
center is $\alpha = 13^h00^m00\hbox{$.\!\!^{s}$}7, \delta =
27^\circ56^\prime51^{\prime\prime}$ (J2000) and $cz_\odot = 7093~\kms$
($cz_{CMB}=7361~\kms$).
The center is 2\farcm3 SW of NGC 4884 and 5\farcm6 ESE of NGC 4874.  
The mass profile agrees with the NFW \citep{nfw97} or
\citet{hernquist1990} form but excludes a singular isothermal sphere. For
the NFW profile, $r_s \simeq 0.17~\Mpc$ and $r_{200} \simeq 1.5~\Mpc$
($r_{200}$ is the radius of the sphere with average mass density 200
times the critical density).  
Varying $q$ changes the mass in the range 6-10$~\Mpc$; the best-fit
analytic form is NFW for $q$=10 and 25 but Hernquist for $q$=50.
Unless otherwise stated, we use $q$=25 in later analysis (this choice
yields the largest mass in the range 6-10$~\Mpc$).  GDK show that the
mass profile agrees with independent X-ray mass estimates 
\citep{hughes89}.

\section{Mass-to-Light Profile}

We subtract background and foreground galaxies (those outside the
caustics) from the sample. 
K01 and \citet{twomdflfn} use 2MASS to calculate the infrared field
galaxy luminosity function (LF) and obtain nearly identical results.
We adopt the values $M^*_{K_s} = -23.39\pm0.05$ and $\alpha = -1.09
\pm 0.06$ (K01) for 2MASS isophotal magnitudes.  Measurements of the
LF of the Coma 
cluster yield similar values of $M^*_{K_s}$ \citep{mt98,dpes98,ap00} and
possibly a steeper faint-end slope ($\alpha \simeq -1.4 \pm 0.3$,$-0.8
\pm 0.4$ and $-1.3 \pm 0.3$ respectively).  At the distance of Coma,
our sample extends 1.2 magnitudes fainter than $M^*_{K_s}$ and includes 
$\approx 68\%$ of the total galaxy light.  This fraction
decreases to 56\% if we adopt $\alpha =-1.3$.  We assume the
luminosity in faint galaxies traces that of the brighter galaxies.  We
estimate the light profile by summing up the luminosity in the bright
member galaxies and multiplying by 1.47 to account for the luminosity
contained in galaxies fainter than $K_s =12.2$.

\begin{figure*}
\figurenum{2}
\label{ml}
%\epsscale{0.4}
\plotone{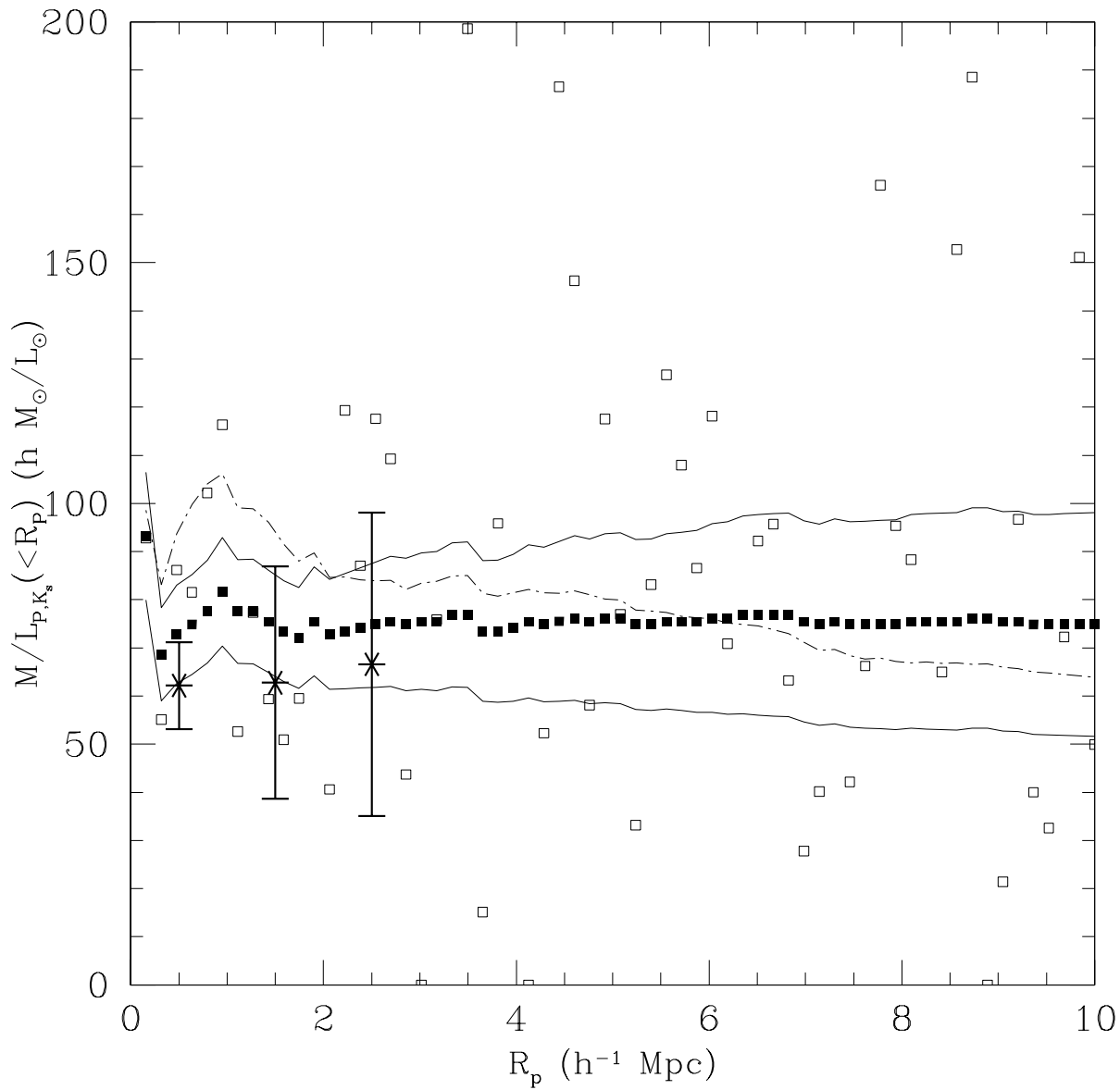} 
\caption{Filled squares indicate the enclosed mass-to-light ratio as
a function of radius; the solid lines show $1-\sigma$ uncertainties.
Open squares show the mass-to-light ratio in 
each spherical shell. The dash-dot line indicates the best-fit
projected Hernquist mass profile divided by the (projected) light
profile.  Stars show the mass-to-light profile from X-ray mass
estimates \citep{hughes89}.}  
\end{figure*}

The resulting $K_s$ band enclosed mass-to-light ratio is constant within
$10~\Mpc$ (Figure \ref{ml}), where $M/L_{K_s}=\mlkedge\pm 23 h$.  We
estimate that the light profile is 
uncertain by $\approx$10\% due to zero points, isophotal magnitudes,
and the LF correction for faint galaxies.  Systematic uncertainties in
the determination of the LF could contribute additional uncertainty
\citep[e.g.,][]{twomdflfn, wright01}.  X-ray mass 
estimates \citep{hughes89} yield similar mass-to-light ratios
(67$\pm32h$ within 2.5$~\Mpc$) and show
no radial trends.  Table \ref{mlvar} lists the
mass-to-light ratios inside and outside $1.6~\Mpc \approx r_{200}$ for
all choices of $q$.  

The light profile is projected; the mass profile is a radial
profile.  Figure \ref{ml} shows the best-fit projected Hernquist mass
profile divided by the projected light profile.  Although an NFW
profile yields a better fit to the mass profile (for $q$=10, 25), the
Hernquist profile is more centrally concentrated and thus shows an
upper bound on this effect (assuming spherical symmetry).  This
profile shows that the mass-to-light ratio may decrease 
with radius.   The mass-to-light ratio inside  $r_{200}$ is at most a
factor of $\sim 2.5$ larger than that outside $r_{200}$ (e.g., 87/36
for $q$=10 in Table \ref{mlvar}).

\section{Discussion}

The K-band mass-to-light ratio of the Coma cluster within $10~\Mpc$ is
$\mlkedge\pm 23 h$ as estimated from the light contained in galaxies
brighter than $K_s=12.2$ and the LF of K01.  We make no correction for
the $\sim$15\% flux omitted by isophotal magnitudes.  A steeper faint-end
slope of $\alpha=-1.3$ would reduce the ratio to $62 \pm 19 h$.
Assuming a typical galaxy color of $B-K\sim 3.7$ \citep{j00} and
$(B-K)_\odot = 2.11$, we obtain $M/L_B \approx 329 \pm 103
h$, in agreement with $M/L_B = 280 \pm 40 h$ from weak lensing on a
similar scale \citep{kaiserxx}.  At a radius of $3^\circ$, $M/L_B
\approx 316 \pm 57 h$; in agreement with \citet{kg82}, who find $M/L_B
\sim 362 h$ at this radius.  X-ray mass estimates yield estimates of 
$280-380 h$ for a mass-follows-light model \citep{hughes89}.
We use a typical galaxy color of $R-K\sim 2.2$ and
$(R-K)_\odot = 0.94$ to estimate  $M/L_R \approx 243\pm 72 h$, in 
agreement with caustic estimates at large radii in A576 (R00). 
We estimate $M/L_H \approx 91\pm27h$, in agreement with
$M/L_H=82-127h$ in A1644 \citep{tustin}.  

The shape of the enclosed mass-to-light profile differs from the
one measured in R band for A576.  Instead of decreasing by a factor of
two between the core and a radius of $4~\Mpc$, the enclosed
mass-to-light ratio is constant within $10~\Mpc$. 
We propose two explanations of this difference.
First, it may be a result of projection effects
\citep{diaferio1999}; the shape of the mass-to-light profile may be
affected by departures from spherical symmetry (R00).  Second,
if the K-band enclosed mass-to-light profile is flat in A576 as in
Coma, a decrease in 
$R-K$ with radius leads to a decreasing profile in R-band.
We expect such a trend if the star formation rate increases with
radius as observed in other systems \citep[e.g.][]{bnm2000}. 
Infrared light profiles 
should be insensitive to recent star formation and best represent the
distribution of stellar mass within the infall region.  
CCD R (K) photometry for Coma (A576) would resolve this issue.
The color gradient effect becomes more significant at bluer
wavelengths; we expect more steeply 
decreasing mass-to-light ratios with decreasing wavelength
\citep[see][]{diaferio1999,bahcall2000}. 

In calculating the mass-to-light profile for Coma, we assume that the
LF is independent of radius; changes in the LF with radius would
affect the mass-to-light profile of Coma.  \citet{bczz01} find
different LFs in field, group, and cluster environments.  Our survey
includes 67, 69, and 65\% of the total light in their field, group, and
cluster LFs respectively assuming a Schechter form.  Thus,
uncertainties due to changes in the LF with environment contribute
$\lesssim$ 7\% uncertainty to the light profile.
Our data provide no constraints
on galaxies fainter than $M^*_{K_s}$-1.2. 

\section{Summary}

We calculate the mass-to-light ratio as a function of radius in the
near-infrared $K_s$ band for the Coma cluster.  This calculation is one of
the first measurements of a cluster mass-to-light ratio in the infrared.
The mass-to-light profile extends to 10~$\Mpc$ and represents one of
the largest scale measurements of a cluster mass-to-light ratio at any
wavelength.  Within 10~$\Mpc$, the enclosed mass-to-light ratio is
of $M/L_K = \mlkedge\pm 23 h$.  With appropriate color
transformations, this value agrees with previous optical and X-ray
estimates for Coma \citep[][]{kg82,hughes89} and estimates at scales of
1-6$\Mpc$ from infall mass estimates (R00) and weak lensing
\citep{kaiserxx} in other systems. 

The enclosed mass-to-light ratio is constant on scales up to
$10~\Mpc$.  This result implies that K-band light measured in bright
galaxies traces the underlying mass distribution in clusters on scales
of up to $10~\Mpc$.  Uncertainties in the mass profile imply that the
mass-to-light ratio inside $r_{200}$ may be as much as a 
factor of $\sim 2.5$ larger than the ratio outside $r_{200}$, possibly
due to antibias.  K-band light is not positively biased with
respect to mass; we cannot rule out antibias.  Radial gradients in the
star formation rate  should create stronger observed antibias at
shorter wavelengths \citep[][R00]{kk1999,bahcall2000,benson00}. 

The asymptotic value of $M/L_{K_s} = \mlkedge\pm 23 h$ implies $\Omega_m =
0.17 \pm 0.05$ using the K01 field galaxy luminosity function.
Because we calculate magnitudes in the same manner as K01 from similar
data, many potential systematic effects should affect our sample and
the field galaxy luminosity function equally.  A recent study of
variations in the luminosity function with environment \citep{bczz01}
suggests that environmental effects contribute $\lesssim$ 7\%
uncertainty to the light profile.  Estimates of possible variations
in $M/L_{K_s}$ with radius (Table \ref{mlvar}) suggest that the density
parameter is no smaller than $\Omega_m \approx 0.08$.  Similar studies
of more distant clusters can produce better constraints
if combined with weak lensing estimates. 

\acknowledgements

This project would not have been possible without the 2MASS team (in
particular M.~Skrutskie, T.~Chester, R.~Cutri, J.~Mader,
and S.E.~Schneider)  or the assistance of
P.~Berlind and M.~Calkins, the remote observers at FLWO and
S.~Tokarz, who processed the spectroscopic data. KR, MJG, MJK, and JPH
are supported in part by the Smithsonian Institution.  AD was
supported by an MPA guest post-doctoral fellowship when this work
began.  This publication makes use of data products from 2MASS, a
joint project of the University of Massachusetts and the Infrared
Processing and Analysis Center, funded by by NASA and NSF.

\bibliography{rines}

\begin{thebibliography}{40}
\expandafter\ifx\csname natexlab\endcsname\relax\def\natexlab#1{#1}\fi

\bibitem[{{Andreon} \& {Pell{\' o}}(2000)}]{ap00}
{Andreon}, S. \& {Pell{\' o}}, R. 2000, \aap, 353, 479

\bibitem[{{Bahcall} {et~al.}(2000){Bahcall}, {Cen}, {Dav{\' e}}, {Ostriker}, \&
  {Yu}}]{bahcall2000}
{Bahcall}, N.~A., {Cen}, R., {Dav{\' e}}, R., {Ostriker}, J.~P., \& {Yu}, Q.
  2000, \apj, 541, 1

\bibitem[{{Balogh} {et~al.}(2001){Balogh}, {Christlein}, {Zabludoff}, \&
  {Zaritsky}}]{bczz01}
{Balogh}, M.~L., {Christlein}, D., {Zabludoff}, A.~I., \& {Zaritsky}, D. 2001,
  \apj, 557, 117

\bibitem[{{Balogh} {et~al.}(2000){Balogh}, {Navarro}, \& {Morris}}]{bnm2000}
{Balogh}, M.~L., {Navarro}, J.~F., \& {Morris}, S.~L. 2000, \apj, 540, 113

\bibitem[{{Benson} {et~al.}(2000){Benson}, {Cole}, {Frenk}, {Baugh}, \&
  {Lacey}}]{benson00}
{Benson}, A.~J., {Cole}, S., {Frenk}, C.~S., {Baugh}, C.~M., \& {Lacey}, C.~G.
  2000, \mnras, 311, 793

\bibitem[{{Carlberg} {et~al.}(1997){Carlberg}, {Yee}, \& {Ellingson}}]{cye97}
{Carlberg}, R.~G., {Yee}, H.~K.~C., \& {Ellingson}, E. 1997, \apj, 478, 462

\bibitem[{{Castander} {et~al.}(2001)}]{sdsscoma}
{Castander}, F.~J. {et~al.} 2001, \aj, 121, 2331

\bibitem[{{Cole} {et~al.}(2001)}]{twomdflfn}
{Cole}, S. {et~al.} 2001, \mnras, 326, 255

\bibitem[{{Colless} \& {Dunn}(1996)}]{cd96}
{Colless}, M. \& {Dunn}, A.~M. 1996, \apj, 458, 435

\bibitem[{{de Propris} {et~al.}(1998){de Propris}, {Eisenhardt}, {Stanford}, \&
  {Dickinson}}]{dpes98}
{de Propris}, R., {Eisenhardt}, P.~R., {Stanford}, S.~A., \& {Dickinson}, M.
  1998, \apjl, 503, L45

\bibitem[{{Diaferio}(1999)}]{diaferio1999}
{Diaferio}, A. 1999, \mnras, 309, 610

\bibitem[{{Diaferio} \& {Geller}(1997)}]{dg97}
{Diaferio}, A. \& {Geller}, M.~J. 1997, \apj, 481, 633

\bibitem[{{Drinkwater} {et~al.}(2001){Drinkwater}, {Gregg}, \&
  {Colless}}]{drink}
{Drinkwater}, M.~J., {Gregg}, M.~D., \& {Colless}, M. 2001, \apj, 548, L139

\bibitem[{{Eisenstein} {et~al.}(1997){Eisenstein}, {Loeb}, \& {Turner}}]{elt97}
{Eisenstein}, D.~J., {Loeb}, A., \& {Turner}, E.~L. 1997, \apj, 475, 421

\bibitem[{{Fabricant} {et~al.}(1998){Fabricant}, {Cheimets}, {Caldwell}, \&
  {Geary}}]{fast}
{Fabricant}, D., {Cheimets}, P., {Caldwell}, N., \& {Geary}, J. 1998, \pasp,
  110, 79

\bibitem[{{Falco} {et~al.}(1999)}]{falco99}
{Falco}, E.~E. {et~al.} 1999, \pasp, 111, 438

\bibitem[{{Gavazzi} \& {Boselli}(1996)}]{gb96}
{Gavazzi}, G. \& {Boselli}, A. 1996, Astrophysical Letters Communications, 35,
  1

\bibitem[{{Gavazzi} {et~al.}(1996){Gavazzi}, {Pierini}, \& {Boselli}}]{gpb96}
{Gavazzi}, G., {Pierini}, D., \& {Boselli}, A. 1996, \aap, 312, 397

\bibitem[{{Geller} {et~al.}(1999){Geller}, {Diaferio}, \& {Kurtz}}]{gdk99}
{Geller}, M.~J., {Diaferio}, A., \& {Kurtz}, M.~J. 1999, \apjl, 517, L23

\bibitem[{{Girardi} {et~al.}(2000){Girardi}, {Borgani}, {Giuricin},
  {Mardirossian}, \& {Mezzetti}}]{g2000}
{Girardi}, M., {Borgani}, S., {Giuricin}, G., {Mardirossian}, F., \&
  {Mezzetti}, M. 2000, \apj, 530, 62

\bibitem[{{Hernquist}(1990)}]{hernquist1990}
{Hernquist}, L. 1990, \apj, 356, 359

\bibitem[{{Hughes}(1989)}]{hughes89}
{Hughes}, J.~P. 1989, \apj, 337, 21

\bibitem[{{Jarrett}(2000)}]{j00}
{Jarrett}, T.~H. 2000, \pasp, 112, 1008

\bibitem[{{Jarrett} {et~al.}(2000){Jarrett}, {Chester}, {Cutri}, {Schneider},
  {Skrutskie}, \& {Huchra}}]{twomassxsc}
{Jarrett}, T.~H., {Chester}, T., {Cutri}, R., {Schneider}, S., {Skrutskie}, M.,
  \& {Huchra}, J.~P. 2000, \aj, 119, 2498

\bibitem[{{Kaiser} {et~al.}(2001){Kaiser}, {Wilson}, {Luppino}, {Kofman},
  {Gioia}, {Metzger}, \& {Dahle}}]{kaiserxx}
{Kaiser}, N., {Wilson}, G., {Luppino}, G., {Kofman}, L., {Gioia}, I.,
  {Metzger}, M., \& {Dahle}, H. 2001, \apj, submitted (astro-ph/9809268)

\bibitem[{{Kent} \& {Gunn}(1982)}]{kg82}
{Kent}, S.~M. \& {Gunn}, J.~E. 1982, \aj, 87, 945

\bibitem[{{Kochanek} {et~al.}(2001)}]{twomasslfn}
{Kochanek}, C.~S. {et~al.} 2001, \apj, submitted (astro-ph/0011456)

\bibitem[{{Kravtsov} \& {Klypin}(1999)}]{kk1999}
{Kravtsov}, A.~V. \& {Klypin}, A.~A. 1999, \apj, 520, 437

\bibitem[{{Mobasher} \& {Trentham}(1998)}]{mt98}
{Mobasher}, B. \& {Trentham}, N. 1998, \mnras, 293, 315

\bibitem[{{Navarro} {et~al.}(1997){Navarro}, {Frenk}, \& {White}}]{nfw97}
{Navarro}, J.~F., {Frenk}, C.~S., \& {White}, S. D.~M. 1997, \apj, 490, 493

\bibitem[{{Nikolaev} {et~al.}(2000){Nikolaev}, {Weinberg}, {Skrutskie},
  {Cutri}, {Wheelock}, {Gizis}, \& {Howard}}]{twomasscalib}
{Nikolaev}, S., {Weinberg}, M.~D., {Skrutskie}, M.~F., {Cutri}, R.~M.,
  {Wheelock}, S.~L., {Gizis}, J.~E., \& {Howard}, E.~M. 2000, \aj, 120, 3340

\bibitem[{{Reisenegger} {et~al.}(2000){Reisenegger}, {Quintana}, {Carrasco}, \&
  {Maze}}]{rqcm}
{Reisenegger}, A., {Quintana}, H., {Carrasco}, E.~R., \& {Maze}, J. 2000, \aj,
  120, 523

\bibitem[{{Rines} {et~al.}(2000){Rines}, {Geller}, {Diaferio}, {Mohr}, \&
  {Wegner}}]{rines2000}
{Rines}, K., {Geller}, M.~J., {Diaferio}, A., {Mohr}, J.~J., \& {Wegner}, G.~A.
  2000, \aj, 120, 2338

\bibitem[{{Skrutskie} {et~al.}(1997)}]{twomass}
{Skrutskie}, M.~F. {et~al.} 1997, in ASSL Vol. 210: The Impact of Large Scale
  Near-IR Sky Surveys, 25--

\bibitem[{{Small} {et~al.}(1998){Small}, {Ma}, {Sargent}, \&
  {Hamilton}}]{small98}
{Small}, T.~A., {Ma}, C., {Sargent}, W. L.~W., \& {Hamilton}, D. 1998, \apj,
  492, 45

\bibitem[{{Tustin} {et~al.}(2001){Tustin}, {Geller}, {Kenyon}, \&
  {Diaferio}}]{tustin}
{Tustin}, A.~W., {Geller}, M.~J., {Kenyon}, S.~J., \& {Diaferio}, A. 2001, \aj,
  in press (astro-ph/0104397)

\bibitem[{{van Haarlem} {et~al.}(1993){van Haarlem}, {Cayon}, {Guiterrez de La
  Cruz}, {Martinez-Gonzalez}, \& {Rebolo}}]{vh93}
{van Haarlem}, M.~P., {Cayon}, L., {Guiterrez de La Cruz}, C.,
  {Martinez-Gonzalez}, E., \& {Rebolo}, R. 1993, \mnras, 264, 71

\bibitem[{{Wegner} {et~al.}(2001)}]{cs2001}
{Wegner}, G. {et~al.} 2001, \aj, in press (astro-ph/0109101)

\bibitem[{{Wright}(2001)}]{wright01}
{Wright}, E.~L. 2001, \apjl, 556, L17

\bibitem[{{Zwicky}(1933)}]{zwicky1933}
{Zwicky}, F. 1933, Helv.~Phys.~Acta, 6, 110

\end{thebibliography}

\begin{table*}[th] \footnotesize
\begin{center}
\caption{\label{mlvar} \sc Radial Variations in $M/L_{K_s}$}
\begin{tabular}{lccccc}
\tableline
\tableline
\tablewidth{0pt}
$q$ & $M_{200}$ & $M(>r_{200})$ & $(M/L_K)_{200}$ & $(M/L_K)(>r_{200})$ & $(M/L_K)(<10~\Mpc)$ \\ 
 & $10^{14}\msun$ & $10^{14}\msun$ & $\mlsun$ & $\mlsun$ & $\mlsun$  \\ 
\tableline
10 & 8.0$\pm$0.6 & 7.7$\pm$1.3 & 71$\pm$9 & 60$\pm$12 & 65$\pm$14 \\
10 H$^a$ & 9.7 & 4.7 & 87 & 36 & 60 \\
25 & 8.1$\pm$0.8 & 10.2$\pm$2.8 & 72$\pm$10 & 79$\pm$23 & \mlkedge$\pm$23 \\
25 H$^a$ & 10.1 & 5.6 & 91 & 43 & 65 \\
50 & 8.5$\pm$1.8 & 6.7$\pm$3.3 & 76$\pm$18 &51$\pm$26 & 63$\pm$35 \\
50 H$^a$ & 10.1 & 5.5 & 91 & 42 & 64 \\
\tableline
$^a${projected best-fit Hernquist profile} \\
\end{tabular}
\end{center}
\end{table*}

\end{document}